\documentclass[twocolumn,pre,superscriptaddress]{revtex4}

\usepackage{graphicx}
\usepackage{amsmath,amssymb}
\usepackage{subfigure}
\usepackage{algorithmic}

\newcommand\etal{\textit{et~al.}}

\newcommand\half{\mbox{$\frac12$}}
\newcommand{\Ord}{\mathrm{O}}
                                   
\begin{document}

\title{An efficient and principled method for detecting communities in
  networks}

\author{Brian Ball}
\affiliation{Department of Physics, University of Michigan, Ann Arbor,
MI 48109, U.S.A.}
\author{Brian Karrer}
\affiliation{Department of Physics, University of Michigan, Ann Arbor,
MI 48109, U.S.A.}
\author{M. E. J. Newman}
\affiliation{Department of Physics, University of Michigan, Ann Arbor,
MI 48109, U.S.A.}
\affiliation{Center for the Study of Complex Systems, University of
  Michigan, Ann Arbor, MI 48109, U.S.A.}

\begin{abstract}
  A fundamental problem in the analysis of network data is the detection of
  network communities, groups of densely interconnected nodes, which may be
  overlapping or disjoint.  Here we describe a method for finding
  overlapping communities based on a principled statistical approach using
  generative network models.  We show how the method can be implemented
  using a fast, closed-form expectation-maximization algorithm that allows
  us to analyze networks of millions of nodes in reasonable running times.
  We test the method both on real-world networks and on synthetic
  benchmarks and find that it gives results competitive with previous
  methods.  We also show that the same approach can be used to extract
  nonoverlapping community divisions via a relaxation method, and
  demonstrate that the algorithm is competitively fast and accurate for the
  nonoverlapping problem.
\end{abstract}

\pacs{89.75.Hc,02.10.Ox,02.50.-r}

\maketitle

\section{Introduction}

Many networked systems, including biological and social networks, are found
to divide naturally into modules or communities, groups of vertices with
relatively dense connections within groups but sparser connections between
them~\cite{GN02,Fortunato10}.  Depending on context, the groups may be
disjoint or overlapping.  A fundamental problem in the theory of networks,
and one that has attracted substantial interest among researchers in the
last decade, is how to detect such communities in empirical network
data~\cite{DDDA05,Fortunato10}.  There are a number of desirable properties
that a good community detection scheme should have.  First, it should be
effective, meaning it should be able to accurately detect community
structure when it is present.  There are, for instance, many examples of
networks, both naturally occurring and synthetic, for which the community
structure is widely agreed upon, and a successful detection method should
be able to find the accepted structure in such cases.  Second, methods
based on sound theoretical principles are preferable over those that are
not.  A method based on a mere hunch that something might work is
inherently less trustworthy than one based on a provable result or
fundamental mathematical insight.  Third, when implemented as a computer
algorithm, a method should ideally be fast and scale well with the size of
the network analyzed.  Many of the networks studied by current science are
large, with millions or even billions of vertices, so a community detection
algorithm whose running time scales, say, linearly with the size of the
network is enormously preferred over one that scales as size squared or
cubed.

In this paper we derive and demonstrate an algorithm for community
detection that can find either overlapping or nonoverlapping communities
and satisfies all of the demands above.  On standard benchmark tests the
algorithm has performance similar to the best previous algorithms in
detecting known community structure.  The algorithm is based on established
methods of statistical inference, namely maximum likelihood and the
expectation-maximization algorithm.  And the algorithm is fast.  In its
simplest form it consists of the iteration of just two sets of equations,
each iteration taking an amount of time that increases only linearly with
system size.  In practice the algorithm can handle networks with millions
of vertices and edges in reasonable running times on a typical desktop
computer.  The largest network we have analyzed has over 4 million vertices
and 40~million edges.

We approach the problem of community detection first as a problem of
finding overlapping communities.  Early efforts at community detection,
going back to the 1970s, assumed nonoverlapping or disjoint communities,
but as many researchers have argued in the last few years, it is common in
practical situations for communities to overlap.  In social networks, for
example, people often belong to more than one circle of
acquaintances---family, friends, coworkers, and so forth---and hence those
circles should properly be considered as overlapping, since they have at
least one common member.  In biological networks too vertices can belong to
more than one group.  Metabolites in a metabolic network can play a role in
more than one metabolic process or cycle; species in a food web can fall on
the border between two otherwise noninteracting subcommunities and play a
role in both of them.  Thus the most general formulation of the community
detection problem should allow for the possibility of overlap.  Our
approach is to develop a solution to this general problem first, then show
how a variant of the same approach can be applied to nonoverlapping
communities as well.

We tackle the detection of overlapping communities by fitting a stochastic
generative model of network structure to observed network data.  This
approach, which applies methods of statistical inference to networks, has
been explored by a number of authors for the nonoverlapping case, including
some work that goes back several
decades~\cite{HLL83,Wang1987,Snijders1997,CMN08}.  Extending the same
approach to the overlapping case, however, has proved nontrivial.  The
crucial step is to devise a generative model that produces networks with
overlapping community structure similar to that seen in real networks.  The
models used in most previous work are ``mixed membership''
models~\cite{Airoldi2008}, in which, typically, vertices can belong to
multiple groups and two vertices are more likely to be connected if they
have more than one group in common.  This, however, implies that the area
of overlap between two communities should have a higher average density of
edges than an area that falls in just a single community.  It is unclear
whether this reflects the behavior of real-world networks accurately, but
it is certainly possible to construct networks that do not have this type
of structure.  Ideally we would prefer a less restrictive model that makes
fewer assumptions about the structure of community overlaps.

Another set of approaches to the detection of overlapping communities are
those based on local community structure.  Rather than splitting an entire
network into communities in one step, these methods instead look for local
groups within the network, based on analysis of local connection patterns
and ignoring global network structure.  Methods of this kind give rise
naturally to overlapping communities when one generates a large number of
independent local communities throughout the network.  Moreover, the
communities tend to be compact and connected subgraphs, a requirement not
always met by other methods.  On the other hand, global detection methods
can capture large-scale network structure better and are more appropriate
when particular constraints, such as constraints on the number of
communities, must be satisfied.

In this paper, we develop a global statistical method for detecting
overlapping communities based on the idea of \emph{link communities} which
has been proposed independently by a number of authors both in the physics
literature~\cite{Ahn2010,Evans2009LG} and in machine
learning~\cite{Parkinnen2009,Gyenge2010}.  The idea is that communities
arise when there are different types of edges in a network.  In a social
network, for instance, there are links representing family ties,
friendship, professional relationships, and so forth.  If we can identify
the types of the edges, i.e.,~if we can cluster not the vertices in a
network but the edges, then we can deduce the communities of vertices after
the fact from the types of edges connected to them.  This approach has the
nice feature of matching our intuitive idea of the origin and nature of
community structure while giving rise to overlapping communities in a
natural way: a vertex belongs to more than one community if it has more
than one type of edge.

Previous approaches to the discovery of link communities have made use of
heuristic quality functions optimized over possible partitions of a
network's edges~\cite{Ahn2010,Evans2009LG}.  Such quality functions,
particularly the so-called modularity function~\cite{NG04}, have been used
in the past for nonoverlapping communities, but while in practice these
functions often give reasonable results, they also have some deficiencies:
the modularity for instance cannot be used to find very small
communities~\cite{FB07}, may not have a unique optimum~\cite{Good2010}, and
is somewhat unsatisfactory from a formal
viewpoint~\cite{Zhang2009,Bickel2009}.  Recent results of Bickel and
Chen~\cite{Bickel2009} suggest that these deficiencies can be remedied by
abandoning the quality function approach and instead fitting a generative
model to the data.  This is the approach we take, but the definition of a
model for link communities demands some subtlety.  In generative models for
\emph{vertex} communities one can assign vertices to groups first and then
place edges based on that assignment.  But for a model of link communities,
where it is the edges that are partitioned, one cannot assign edges to
groups until the edges exist, so the edges and their groupings have to be
generated simultaneously.  We describe in detail how we achieve this in the
following section.  Once we have the model, the goal will be to determine
the values of its parameters that best fit the observed network and from
those to determine the overlapping vertex communities.

The outline of the paper is as follows.  First we define our model and then
demonstrate how the best-fit values of its parameters can be calculated
using a maximum likelihood algorithm.  In its simplest form this algorithm
is only moderately fast, but we demonstrate that many of the model
parameters converge rapidly to trivial values and hence can be pruned from
the calculation.  We give a prescription for performing this pruning,
resulting in a significantly faster algorithm which is practical for
applications to very large networks.  We give example applications to
numerous real-world networks, as well as tests against synthetic networks
that demonstrate that the algorithm can discover known overlapping
community structure in such networks.

Finally, we show how our method can be used also to detect nonoverlapping
communities by assigning each vertex solely to the community to which it
most strongly belongs in the overlapping division.  We demonstrate that
this intuitive heuristic can be justified rigorously by regarding the link
community model as a relaxation of a stochastic blockmodel for disjoint
communities~\cite{Karrer2010SB}.  Algorithms have been proposed previously
for fitting this blockmodel, but their running time was always at least
quadratic in the number of vertices, which limited their application to
smaller networks.  The algorithm proposed here is significantly faster and
hence can be applied to the detection of disjoint communities in very large
networks.

\section{A generative model for link communities}
\label{sec:overlap}
Our first step is to define the generative network model that we will use.
The model generates networks with a given number~$n$ of vertices and
undirected edges divided among a given number~$K$ of communities.  It is
convenient to think of the edges as being colored with $K$ different colors
to represent the communities to which they belong.  Then the model is
parametrized by a set of parameters~$\theta_{iz}$, which represent the
propensity of vertex~$i$ to have edges of color~$z$.  Specifically,
$\theta_{iz}\theta_{jz}$ is the expected number of edges of color~$z$ that
lie between vertices~$i$ and~$j$, the exact number being Poisson
distributed about this mean value.  Note that this means the network is
technically a multigraph---it can have more than one edge between a pair of
vertices.  Most real-world networks have single edges only, and in this
sense the model is unrealistic.  However, allowing multiedges makes the
model enormously simpler to treat and in practice the values of the
$\theta_{iz}$ will be small so that the number of multiedges, and hence the
error introduced, is also small.  The same approximation is made in most
other random graph models of networks, including, for instance, the widely
studied configuration model~\cite{MR95,NSW01}.  Our model also allows
self-edges---edges that connect to the same vertex at both ends---with
expected number $\half\theta_{iz}\theta_{iz}$, the extra factor of a half
being convenient for consistency with later results.  Again, the appearance
of self-edges, while unrealistic in some cases, greatly simplifies the
mathematical developments and is typical of other random graph models
including the configuration model.

In the model defined here the link communities arise implicitly as the
network is generated, as discussed in the introduction, rather than being
spelled out explicitly.  Two vertices~$i,j$ which have large values of
$\theta_{iz}$ and $\theta_{jz}$ for some value of~$z$ have a high
probability of being connected by an edge of color~$z$, and hence groups of
such vertices will tend to be connected by relatively dense webs of
color-$z$ edges---precisely the structure we expect to see in a network
with link communities.

\section{Detecting overlapping communities}
Given the model defined above, it is now straightforward to write down the
probability with which any particular network is generated.  Recalling that
the sum of independent Poisson-distributed random variables is also a
Poisson-distributed random variable, the expected total number of edges of
all colors between two vertices $i$ and~$j$ is simply $\sum_z
\theta_{iz}\theta_{jz}$ (or $\half \sum_z \theta_{iz}\theta_{iz}$ for
self-edges), and the actual number is Poisson-distributed with this mean.
Thus the probability of generating a graph~$G$ with adjacency matrix
elements~$A_{ij}$ is
\begin{align}
P(G|\theta) &= \prod_{i<j}
  {\bigl(\sum_z\theta_{iz}\theta_{jz}\bigr)^{A_{ij}}\over A_{ij}!}
  \exp \Bigl( {-\sum_z\theta_{iz}\theta_{jz}} \Bigr) \nonumber\\
  & \hspace{-1em}{} \times \prod_i
  {\bigl( \half\sum_z \theta_{iz}\theta_{iz} \bigr)^{A_{ii}/2}\over
  (A_{ii}/2)!} \exp\Bigl( {-\half\sum_z \theta_{iz}\theta_{iz}}
  \Bigr).
\label{eq:likelihood}
\end{align}
(Recall that the adjacency matrix element~$A_{ij}$, by convention, takes
the value $A_{ij}=1$ if there is an edge between distinct vertices $i$
and~$j$, but $A_{ii}=2$ for a self-edge---hence the additional factors
of~$\half$ in the second product.)

We fit the model to an observed network by maximizing this probability with
respect to the parameters~$\theta_{iz}$, or equivalently (and more
conveniently) maximizing its logarithm.  Taking the log of
Eq.~\eqref{eq:likelihood}, rearranging, and dropping additive and
multiplicative constants (which have no effect on the position of the
maximum), we derive the log-likelihood
\begin{equation}
\log P(G|\theta) = \sum_{ij} A_{ij}
    \log\bigl( {\textstyle\sum_z \theta_{iz}\theta_{jz}} \bigr)
    - \sum_{ijz} \theta_{iz}\theta_{jz}.
\label{eq:ll}
\end{equation}
Direct maximization of this expression by differentiating leads to a set of
nonlinear implicit equations for~$\theta_{iz}$ that are hard to solve, even
numerically.  An easier approach is the following.  First we apply Jensen's
inequality in the form~\cite{note1}:
\begin{equation}
\log \bigl( {\textstyle\sum_z x_z} \bigr) \ge \sum_z q_z \log {x_z\over
  q_z},
\label{eq:jensen}
\end{equation}
where the $x_z$ are any set of positive numbers and the $q_z$ are any
probabilities satisfying $\sum_z q_z = 1$.  Note that the exact equality
can always be achieved by making the particular choice $q_z=x_z/\sum_z
x_z$.  Applying Eq.~\eqref{eq:jensen} to Eq.~\eqref{eq:ll} gives
\begin{equation}
\log P(G|\theta) \ge \sum_{ijz} \biggl[ A_{ij} q_{ij}(z) \log
  {\theta_{iz}\theta_{jz}\over q_{ij}(z)}  - \theta_{iz}\theta_{jz} \biggr],
\label{eq:ineq}
\end{equation}
where the probabilities~$q_{ij}(z)$ can be chosen in any way we please
provided they satisfy $\sum_z q_{ij}(z) = 1$.  Notice that the $q_{ij}(z)$
are only defined for vertex pairs~$i,j$ that are actually connected by an
edge in the network (so that~$A_{ij}=1$), and hence there are only as many
of them as there are observed edges.

Since, as noted, the exact equality in this expression can always be
achieved by a suitable choice of~$q_{ij}(z)$, it follows that the double
maximization of the right-hand side of~\eqref{eq:ineq} with respect to both
the~$q_{ij}(z)$ and the~$\theta_{iz}$ is equivalent to maximizing the
original log-likelihood, Eq.~\eqref{eq:ll}, with respect to
the~$\theta_{iz}$ alone.  It may appear that this does not make our
optimization problem any simpler: we have succeeded only in turning a
single optimization into a double one, which one might well imagine was a
more difficult problem.  Delightfully, however, it is not;
the double optimization is actually very simple.  Given the true optimal
values of~$\theta_{iz}$, the optimal values of~$q_{ij}(z)$ are given by
\begin{equation}
q_{ij}(z) = \frac{\theta_{iz}\theta_{jz}}{\sum_{z}\theta_{iz}\theta_{jz}},
\label{eq:estep}
\end{equation}
since these are the values that make our inequality an exact equality.  But
given the optimal values of the~$q_{ij}(z)$, the optimal~$\theta_{iz}$ can
be found by differentiating~\eqref{eq:ineq}, which gives
\begin{equation}
\theta_{iz} = \frac{\sum_j A_{ij}q_{ij}(z)}{\sum_i \theta_{iz}}.
\label{eq:mstep0}
\end{equation}
Summing this expression over~$i$ and rearranging gives us
\begin{equation}
\Bigl( \sum_i \theta_{iz} \Bigr)^2 = \sum_{ij} A_{ij} q_{ij}(z),
\end{equation}
and combining with~\eqref{eq:mstep0} again then gives
\begin{equation}
\theta_{iz} = \frac{\sum_j A_{ij}q_{ij}(z)}{\sqrt{\sum_{ij} A_{ij}q_{ij}(z)}}.
\label{eq:mstep}
\end{equation}
Maximizing the log-likelihood is now simply a matter of simultaneously
solving Eqs.~\eqref{eq:estep} and~\eqref{eq:mstep}, which can be done
iteratively by choosing a random set of initial values and alternating back
and forth between the two equations.  This type of approach is known as an
expectation-maximization or EM algorithm and it can be proved that the
log-likelihood increases monotonically under the iteration, though it does
not necessarily converge to the global maximum.  To guard against the
possibility of getting stuck in a local maximum, we repeat the entire
calculation a number of times with random initial conditions and choose the
result that gives the highest final log-likelihood.

The value of $q_{ij}(z)$ in Eq.~\eqref{eq:estep} has a simple physical
interpretation: it is the probability that an edge between $i$ and $j$ has
color~$z$, which is precisely the quantity we need in order to infer link
communities in the network.  Notice that $q_{ij}(z)$ is symmetric in $i,j$,
as it should be for an undirected network.

The calculation presented here is mathematically closely related to methods
developed in the machine learning community for the analysis of text
documents.  Specifically, the model we fit can be regarded as a variant of
a model used in probabilistic latent semantic analysis (PLSA)---a technique
for automated detection of topics in a corpus of text---adapted to the
present context of link communities.  Connections between text analysis and
community detection have been explored by several previous authors.  Of
particular interest is the work of Psorakis~\etal~\cite{Psorakis2010},
which, though it does not focus on link communities, uses another variant
of the PLSA model, coupling it with an iterative fitting algorithm called
nonnegative matrix factorization to find overlapping communities in
directed networks.  Also of note is the work of
Parkinnen~\etal~\cite{Parkinnen2009}, who consider link communities as we
do, but take a contrasting algorithmic approach based on a Bayesian
generative model and Markov chain Monte Carlo techniques.  A detailed
description of the interesting connections between text processing and
network analysis would take us some way from the primary purpose of this
paper, but for the interested reader we give a discussion and references in
Appendix~\ref{app:plsa}.

\section{Implementation}
\label{sec:implementation}
The method outlined above can be implemented directly as a computer
algorithm for finding overlapping communities, and works well for networks
of moderate size, up to tens of thousands of vertices.  For larger networks
both memory usage and run-time become substantial and prevent the
application of the method to the largest systems, but both can be improved
by using a more sophisticated implementation which makes applications to
networks of millions of vertices possible.

The algorithm's memory use is determined by the space required to store the
parameters: the $\theta_{iz}$ require $\Ord(nK)$ space while
the~$q_{ij}(z)$ require~$\Ord(mK)$, where $n$ and $m$ are the numbers of
vertices and edges in the network.  Since $m$ is usually substantially
larger than~$n$, this means that memory use is dominated by
the~$q_{ij}(z)$.  We can reduce memory use by reorganizing the algorithm in
such a way that the~$q_{ij}(z)$ are never stored.  Rather than focusing on
the~$\theta_{iz}$, we work instead with the average number~$k_{iz}$ of ends
of edges of color~$z$ connected to vertex~$i$:
\begin{equation}
k_{iz} = \sum_j A_{ij} q_{ij}(z).
\label{eq:kiziter}
\end{equation}
Given the values of these quantities on a given iteration of the algorithm,
the calculation of the values at the next iteration is then as follows.
First we define a new set of quantities~$k_{iz}'$ that will store the new
values of the~$k_{iz}$.  Initially we set all of them to zero.  We also
calculate the average number~$\kappa_z$ of edges of color~$z$ summed over
all vertices
\begin{equation}
\kappa_z = \sum_i k_{iz},
\label{eq:kappaziter}
\end{equation}
in terms of which the original $\theta_{iz}$ parameters are
\begin{equation}
\theta_{iz} = {k_{iz}\over\sqrt{\kappa_z}}.
\end{equation}
Next we go through each edge~$(i,j)$ in the network in turn and calculate
the denominator of Eq.~\eqref{eq:estep} for that $i$ and~$j$ from the
values of the~$k_{iz}$ thus:
\begin{equation}
D = \sum_z \theta_{iz}\theta_{jz} = \sum_z {k_{iz}k_{jz}\over\kappa_z}.
\end{equation}
Armed with this value we can calculate the value of~$q_{ij}(z)$ for this
$i,j$ and all~$z$ from Eq.~\eqref{eq:estep}:
\begin{equation}
q_{ij}(z) = {\theta_{iz}\theta_{jz}\over\sum_z \theta_{iz}\theta_{jz}}
          = {k_{iz}k_{jz}\over D\kappa_z}.
\end{equation}
Now we add this value onto the quantities~$k_{iz}'$ and $k_{jz}'$, discard
the values of $D$ and~$q_{ij}(z)$, and repeat for the next edge in the
network.  When we have gone through all edges in this manner, the
quantities~$k_{iz}'$ will be equal to the sum in Eq.~\eqref{eq:kiziter},
and hence will be the correct new values of~$k_{iz}$.

This method requires us to store only the old and new values of~$k_{iz}$,
for a total of $2nK$ quantities, and not the values of~$q_{ij}(z)$, which
results in substantial memory savings for larger networks.

As for the running time, the algorithm as we have described it has a
computational complexity of $\Ord(mK)$ operations per iteration of the
equations, where $m$ is again the number of edges in the network.  The
primary determinant of the total run-time is the number of iterations that
have to be performed before the values of the $k_{iz}$ converge.  In
practice, we find in many cases that a rather large number of iterations is
required, which slows the performance of the method, but the speed can be
improved as follows.

In a typical application of the algorithm to a network, the end result is
that each vertex belongs to only a subset of the $K$ possible communities.
To put that another way, we expect that a large fraction of the
parameters~$k_{iz}$ will tend to zero under the EM iteration.  It is
straightforward to see from the equations above that if a
particular~$k_{iz}$ ever becomes zero, then it must remain so for all
future iterations, which means that it no longer need be updated and we can
save ourselves time by excluding it from our calculations.  This leads to
two useful strategies for pruning our set of variables.  In the first, we
set to zero any $k_{iz}$ that falls below a predetermined
threshold~$\delta$.  Once a $k_{iz}$ has been set to zero, the
corresponding values of the~$q_{ij}(z)$ on all the adjacent edges are also
zero and therefore need not be calculated.  Thus, for each edge, we need
only calculate the values of~$q_{ij}(z)$ for those colors~$z$ for which
both~$k_{iz}$ and $k_{jz}$ are nonzero, i.e.,~for the intersection of the
sets of colors at vertices~$i$ and~$j$.  This strategy leads to significant
speed increases when the number~$K$ of communities is large.  For smaller
values of~$K$ the speed savings are outweighed by the additional
computational overhead and it is more efficient to simply calculate
all~$q_{ij}(z)$, but we nonetheless still set the values of the~$k_{iz}$ to
zero below the threshold~$\delta$ because it makes possible our second
pruning strategy.

Our second strategy, which can be used in tandem with the first and gives
significant speed improvements for all values of~$K$, is motivated by the
observation that if all but one of the $k_{iz}$ for a particular vertex are
set to zero, then the color of the vertex is fixed at a single value and
will no longer change at all.  If both vertices at the ends of an
edge~$(i,j)$ have this property, if both of them have converged to a single
color and are no longer changing, then the edge connecting them no longer
has any effect on the calculation and can be deleted entirely from the
network.

By the use of these two strategies the speed of our calculations is
improved markedly.  We find in practice that the numbers of
parameters~$k_{iz}$ and edges both shrink rapidly and substantially with
the progress of the calculation, so that the majority of the iterations
involve only a subset, typically those associated with the vertices whose
community identification is most ambiguous.  If the value of the
threshold~$\delta$ is set to zero, then the pruned algorithm is exactly
equivalent to the original EM algorithm and the results are identical, yet
even with this choice we find substantial speed improvements.  If $\delta$
is chosen small but nonzero---we use $\delta=0.001$ in our
calculations~\cite{note2}---then we introduce an approximation into the
calculation which means the results will be different in general from the
original algorithm.  In practice, however, the difference is small, and the
nonzero~$\delta$ gives us an additional significant speed improvement.

A detailed comparison of results and run-times for the pruned and original
versions of the algorithm is given in Appendix~\ref{app:alg} for a range of
networks.  Unless stated otherwise, all calculations
presented in the remainder of the paper are done with the faster version of
the algorithm.

\begin{figure*}
\begin{center}
\includegraphics[height=4.2cm,clip=true]{edge_color_av_degree.permdf_v4.eps}
\hfill
\includegraphics[height=4.2cm,clip=true]{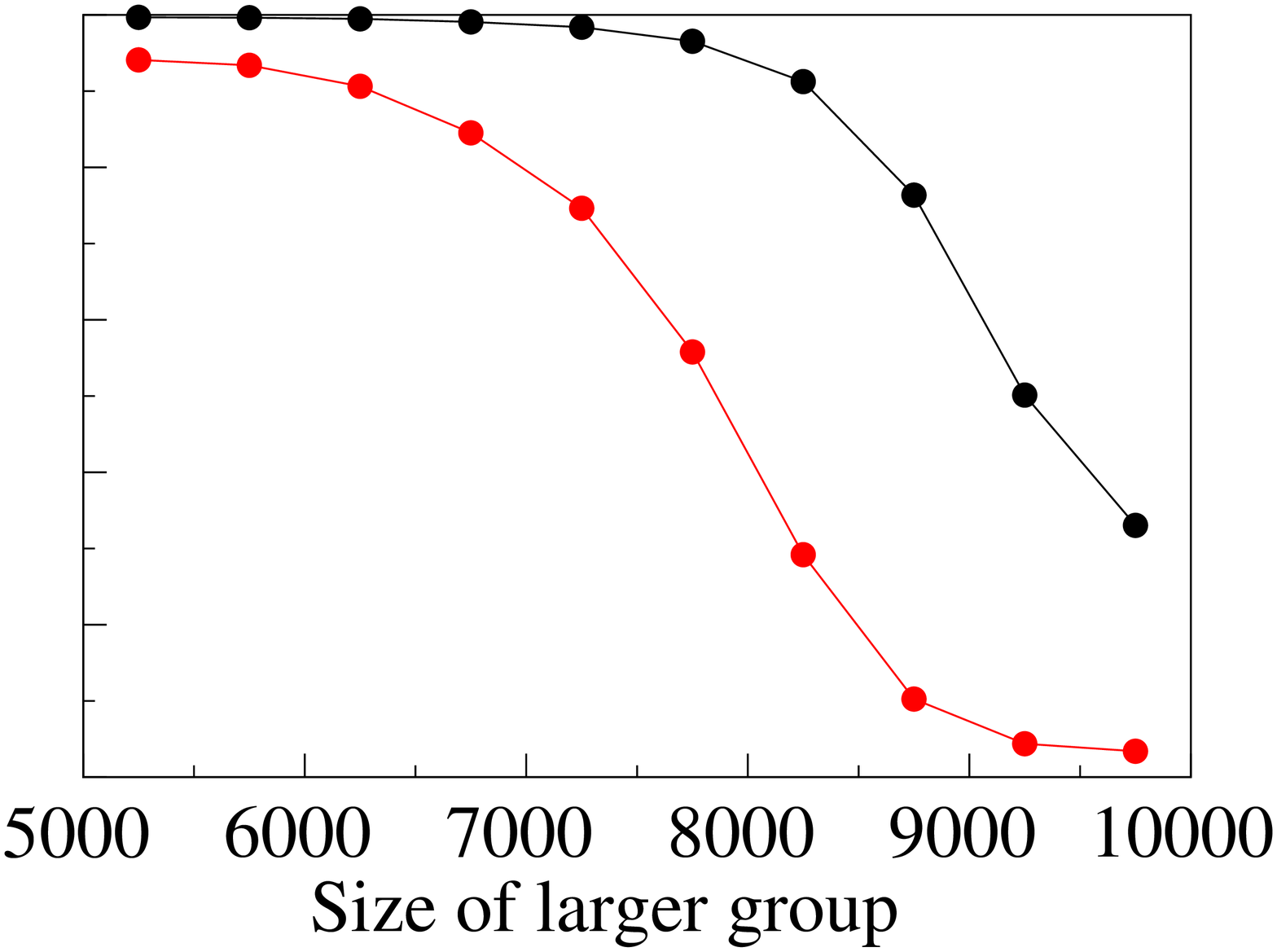}
\hfill
\includegraphics[height=4.2cm,clip=true]{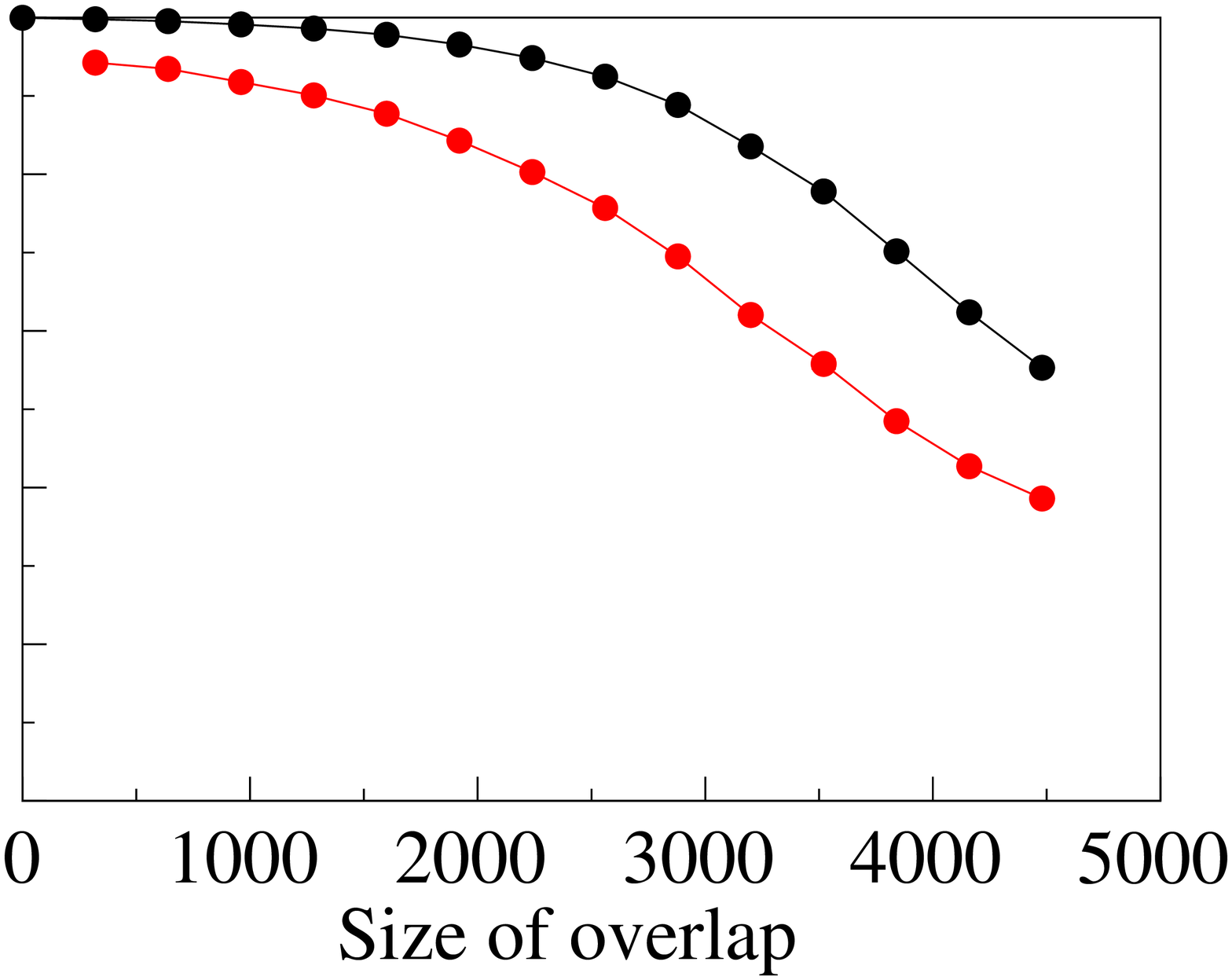}
\end{center}
\caption{Results from the three sets of synthetic tests described in the
  text.  Each data point is averaged over 100 networks.  Twenty random
  initializations of the variables were used for each network and the run
  giving the highest value of the log-likelihood was taken as the final
  result.  In each panel the black curve shows the fraction of vertices
  assigned to the correct communities by the algorithm, while the red curve
  is the Jaccard index for the vertices in the overlap.}
\label{fig:othertests}
\end{figure*}

\section{Results}
We test the performance of the algorithm described above using both
synthetic (computer-generated) networks and a range of real-world examples.
The synthetic networks allow us to test the algorithm's ability to detect
known, planted community structure under controlled conditions, while the
real networks allow us to observe performance under practical, real-world
conditions.

\subsection{Synthetic networks}
Our synthetic network examples take the form of a classic consistency test.
We generate networks using the same stochastic model that the algorithm
itself is based on and measure the algorithm's ability to recover the known
community divisions for various values of the parameters.  One can vary the
values to create networks with stark community structure (which should make
detection easy) or no community structure at all (which makes it
impossible), and everything in between, and we can thereby vary the
difficulty of the challenge we pose to the algorithm.

The networks we use for our tests have $n=10000$ vertices each, divided
into two overlapping communities.  We place $x$ vertices in the first
community only, meaning they have connections only to others in that
community, $y$~vertices in the second community only, and the remaining
$z=n-x-y$ vertices in both communities, with equal numbers of connections
to vertices in either group on average.  We fix the expected degree of all
vertices to take the same value~$k$.

We perform three sets of tests.  In the first we fix the size of the
overlap between the communities at $z=500$, divide the remaining vertices
evenly $x=y=4750$, and observe the behavior of the algorithm as we vary the
value of~$k$.  When $k\to0$ there are no edges in the network and hence no
community structure, and we expect the algorithm (or any algorithm) to
fail.  When $k$ is large, on the other hand, it should be straightforward
to work out where the communities are.

For our second set of tests we again set the overlap at $z=500$ but this
time we fix $k=10$ and vary the balance of vertices between $x$ and~$y$.
Finally, for our third set of tests we set $k=10$ and constrain $x$ and $y$
to be equal, but allow the size~$z$ of the overlap to vary.

In Fig.~\ref{fig:othertests} we show the measured fraction of vertices
classified correctly (black curve) in each of these three sets of tests
(the three separate panels), averaged over 100 networks for each point.  To
be considered correctly classified a vertex's membership (or lack of
membership) in both groups must be reported correctly by the algorithm, and
the algorithm considers any vertex to be a member of a group if, on
average, it has at least one edge of the appropriate color when the maximum
likelihood fitting procedure is complete.  In mathematical terms, a vertex
belongs to community~$z$ if its expected degree with respect to color~$z$,
given by $\sum_j A_{ij} q_{ij}(z)$, is greater than one.

As the figure shows, there are substantial parameter ranges for all three
tests for which the algorithm performs well, correctly classifying most of
the vertices in the network.  As expected the accuracy in the first test
increases with increasing~$k$ and for values of~$k$ greater than about
ten---a figure easily attained by many real-world networks---the algorithm
identifies the known community structure essentially perfectly.  In the
other two tests accuracy declines as either the asymmetry of the two groups
or the size of the overlap increases, but approaches 100\% when either is
small.

To probe in more detail the algorithm's ability to identify overlapping
communities, we have also measured, for the same test networks, a Jaccard
index: if $S$ is the set of vertices in the true overlap and $V$ is the set
the algorithm identifies as being in the overlap, then the Jaccard index is
$J=|S\cap V|/|S\cup V|$.  This index is a standard measure of similarity
between sets that rewards accurate identification of the overlap while
penalizing both false positives and false negatives.  The values of the
index are shown as the red curves in Fig.~\ref{fig:othertests} and, as we
can see, the general shape of the curves is similar to the overall fraction
of correctly identified vertices.  In particular, we note that for networks
with sufficiently high average degree~$k$ the value of~$J$ tends to~1,
implying that the overlap is identified essentially perfectly.

\subsection{Real networks}
We have also tested our method on numerous real-world networks.  In this
section we give detailed results for three specific examples.  Summary
results for a number of additional examples are given in
Appendix~\ref{app:alg}.

Our first example is one that has become virtually obligatory in tests of
community detection, Zachary's ``karate club'' network, which represents
friendship patterns between members of a university sports club, deduced
from an observational study~\cite{Zachary77}.  The network is interesting
because the club split in two during the study, as a result of an internal
dispute, and it has been found repeatedly that one can deduce the lines of
the split from a knowledge of the network structure alone.

\begin{figure}
\begin{center}
\subfigure[]{\includegraphics[width=8cm]{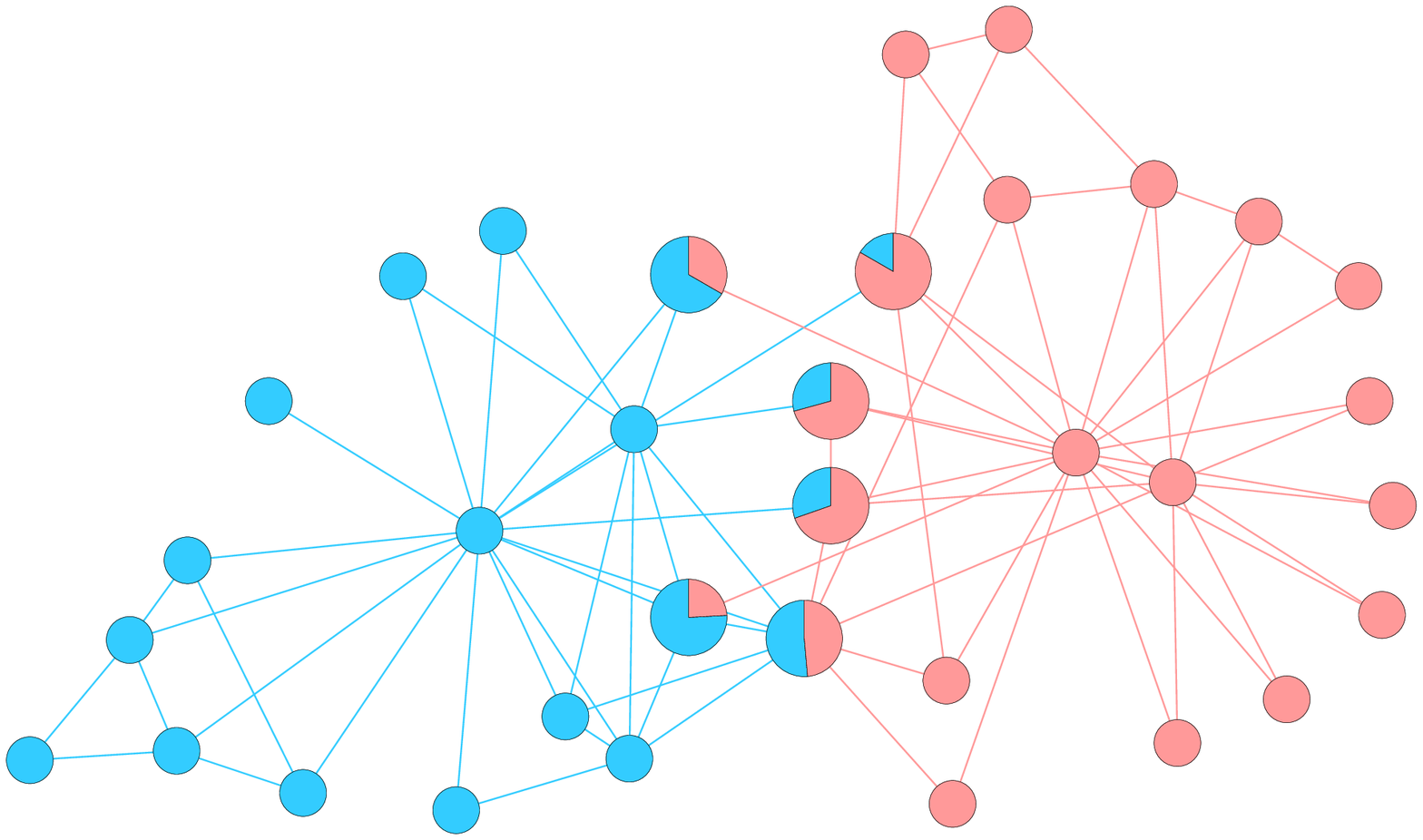}}
\subfigure[]{\includegraphics[width=8cm]{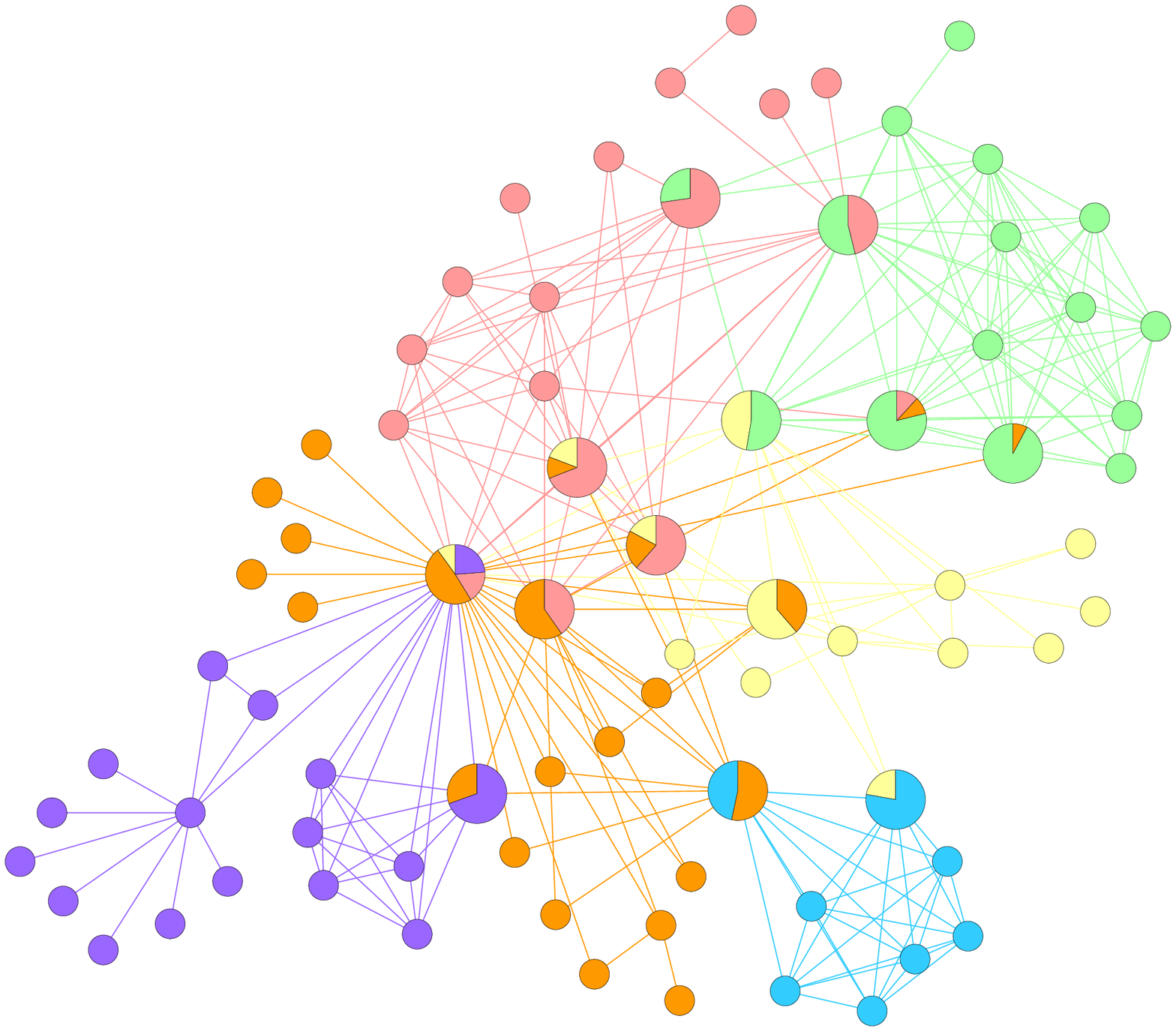}}
\end{center}
\caption{Overlapping communities in (a)~the karate club network
  of~\cite{Zachary77} and (b)~the network of characters from \textit{Les
    Mis\'erables}~\cite{Knuth93}, as calculated using the algorithm
  described in this paper.  The edge colors correspond to the highest value
  of~$q_{ij}(z)$ for the given edge, while vertex colors indicate the
  fraction of incident edges that fall in each community.  For vertices in
  more than one community the vertices are drawn larger for clarity and
  divided into pie charts representing their division among communities.}
\label{fig:combination}
\end{figure}

Figure~\ref{fig:combination}a shows the decomposition of the karate club
network into two overlapping groups as found by our algorithm.  The colors
in the figure show both the division of the vertices and the division of
the edges.  The split between the two groups in the club is clearly evident
in the results and corresponds well with the acknowledged ``ground truth,''
but in addition the algorithm assigns several vertices to both groups.  The
individuals represented by these overlap vertices, being by definition
those who have friends in both camps, might be supposed to have had some
difficulty deciding which side of the dispute to come down on, and indeed
Zachary's original discussion of the split includes some indications that
this was the case.  Note also that, in addition to identifying overlapping
vertices, our method can assign to each a fraction by which it belongs to
one community or the other, represented in the figure by the pie-chart
coloring of the vertices in the overlap.  The fraction is calculated as the
expected fraction of edges of each color incident on the vertex.

Our second example is another social network and again one whose community
structure has been studied previously.  This network, compiled by
Knuth~\cite{Knuth93}, represents the patterns of interactions between the
fictional characters in the novel \textit{Les Mis\'erables} by Victor Hugo.
In this network two characters are connected by an edge if they appear in
the same chapter of the book.  Figure~\ref{fig:combination}b shows our
algorithm's partition of the network into six overlapping communities and
the partition accords roughly with social divisions and subplots in the
plot-line of the novel.  But what is particularly interesting in this case
is the role played by the hubs in the network---the major characters who
are represented by vertices of especially high degree.  It is common to
find high-degree hubs in networks of many kinds, vertices with so many
connections that they have links to every part of the network, and their
presence causes problems for traditional, nonoverlapping community
detection schemes because they do not fit comfortably in any community: no
matter where we place a hub it is going to have many connections to
vertices in other communities.  Overlapping communities provide an elegant
solution to this problem because we can place the hubs in the overlaps.  As
Fig.~\ref{fig:combination}b shows, our algorithm does exactly this, placing
many of the hubs in the network in two or more communities.  Such an
assignment is in this case also realistic in terms of the plot of the
novel: the major characters represented by the hubs are precisely those
that appear in more than one of the book's subplots.

\begin{figure*}
\begin{center}
\includegraphics[width=17.5cm]{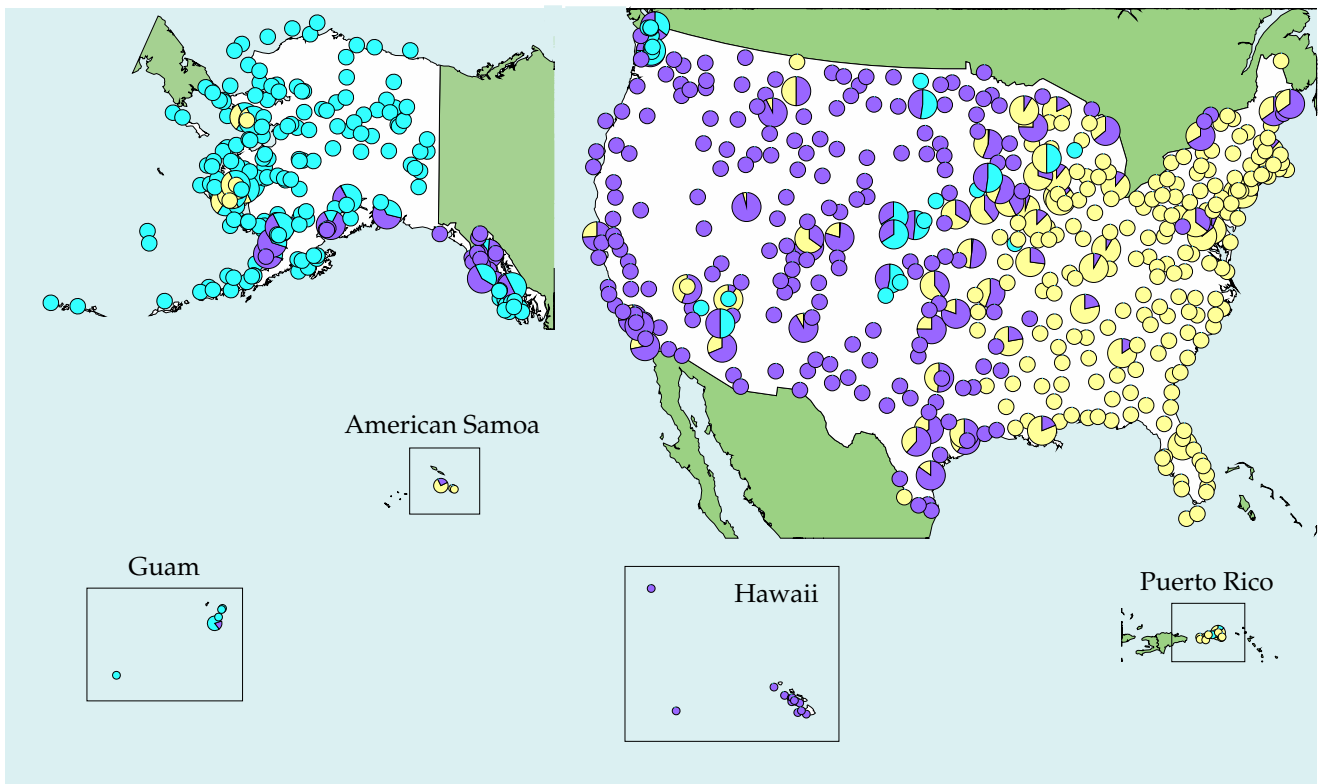}
\end{center}
\caption{Overlapping communities in the network of US passenger air
  transportation.  The three communities produced by the calculation
  correspond roughly to the east and west coasts of the country and
  Alaska.}
\label{airgraph}
\end{figure*}  

A similar behavior can be seen in our third example, which is a
transportation network, the network of passenger airline flights between
airports in the United States.  In this network, based on data for flights
in 2004, the vertices represent airports and an edge between airports
indicates a regular scheduled direct flight.  Spatial networks, those in
which, as here, the vertices have well-defined positions in geographic
space, are often found to have higher probability of connection for vertex
pairs located closer together~\cite{GN06b,Barthelemy2010}, which suggests
that communities, if they exist, should be regional, consisting principally
of blocks of nearby vertices.  The communities detected by our algorithm in
the airline network follow this pattern, as shown in Fig.~\ref{airgraph}.
The three-way split shown divides the network into east and west coast
groups and a group for Alaska.  The overlaps are composed partly of
vertices that lie along the geographic boundaries between the groups, but
again include hubs as well, which tend to be placed in the overlaps even
when they don't lie on boundaries.  As with the previous example, this
placement gives the algorithm a solution to the otherwise difficult problem
of assigning to any one group a hub that has connections to all parts of
the network, but it also makes intuitive sense.  Hubs are the ``brokers''
of the airline network, the vertices that connect different communities
together, since they are precisely the airports that one passes through in
traveling between distant locations.  Thus it is appropriate that they be
considered members of more than one group.  In most cases the hubs belong
most strongly to the community in which they are geographically located,
and less strongly to other communities.

\section{Nonoverlapping communities}
\label{sec:nonoverlap}

As we have described it, our algorithm is an algorithm for finding
overlapping communities in networks, but it can be used to find
nonoverlapping communities as well.  As pointed out by a number of previous
authors~\cite{Zarei2009,Wang2010,Psorakis2010}, any algorithm that
calculates proportional membership of vertices in communities can be
adapted to the nonoverlapping case by assigning each vertex to the single
community to which it belongs most strongly.  In our case, this means
assigning vertices to the community for which the value of~$\theta_{iz}$ is
largest.  It turns out that this procedure can be justified rigorously in
our case by regarding the link community model as a relaxation of a
nonoverlapping degree-corrected stochastic blockmodel.  The details are
given in Appendix~\ref{app:relaxation}.  Here we give some example
applications to show how the approach works in practice.

As with the overlapping case, we test the method on both synthetic and
real-world networks.  For the synthetic case we use a standard test, the
LFR benchmark for unweighted undirected networks with planted community
structure~\cite{Lancichinetti2008,Lancichinetti2009}.  To make comparisons
simple we use the same parameters as in Ref.~\cite{Lancichinetti2009} with
networks of 1000 and 5000 vertices, average degree~20, maximum degree~50,
degree exponent~$-2$, and community exponent~$-1$.  We also use the same
two ranges of community sizes, with communities of 10 to 50 vertices for
one set of tests (labeled S for ``small'' in our figures) and 20 to 100
vertices for the other set (labeled B for ``big'').  The value of $K$ for
the detection algorithm was set equal to the number of communities in the
benchmark network (which, because of the nature of the benchmark, is not a
constant but varies from one network to another).

To quantify our algorithm's success at detecting the known communities in
the benchmark networks we use the variant normalized mutual information
measure proposed in~\cite{Lancichinetti2009}.  We note that this measure is
different, and in general returns different results, from the traditional
normalized mutual information used to evaluate community
structure~\cite{DDDA05}, but using it allows us to make direct comparisons
with the results for other algorithms given in~\cite{Lancichinetti2009}.

In our benchmark tests we find that the simplistic rounding method
described above for nonoverlapping communities, just choosing the community
with the highest value of~$\theta_{iz}$, returns only average performance
when compared to the other algorithms tested in
Ref.~\cite{Lancichinetti2009}.  However, a simple modification of the
algorithm produces significantly better results: after generating a
candidate division into communities using the rounding method, we then
apply a further optimization step in which we move from one community to
another the single vertex that most increases the log-likelihood of the
division under the stochastic blockmodel, and repeat this exercise until no
further such moves exist.  This process, which is easy to implement and
carries little computational cost when compared to the calculation of
the initial division, improves our results dramatically.

\begin{figure}
\begin{center}
\includegraphics[width=8cm]{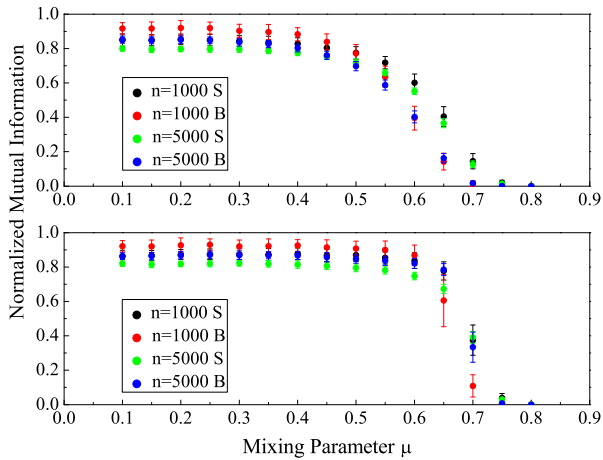}
\end{center}
\caption{Results of tests of the nonoverlapping community algorithm
  described in the text applied to synthetic networks generated using the
  LFR benckmark model of Lancichinetti~\etal~\cite{Lancichinetti2009}.
  Parameters used were the same as in Ref.~\cite{Lancichinetti2009} and (S)
  and (B) denote the ``small'' and ``big'' community sizes used by the same
  authors.  The top and bottom panels respectively show the results without
  and with post-processing to optimize the value of the log-likelihood.
  Ten random initializations of the variables were used for each network
  and each point is an average over 100 networks.}
\label{fig:LFRBench}
\end{figure}

The results of our tests are shown in Figure~\ref{fig:LFRBench}.  The top
panel shows the performance of the algorithm without the additional
optimization step and the results fall in the middle of the pack when
compared to previous algorithms, better than some methods but not as good
as others.  The bottom panel shows the results with the additional
optimization step added, and now the algorithm performs about as well as,
or better than, the algorithms analyzed in Ref.~\cite{Lancichinetti2009}.
The general shape of the mutual information curve is similar to that of the
best competing methods, falling off around the same place, although the
mutual information values are somewhat lower for low values of the mixing
parameter, indicating that the method is not getting the community
structure exactly correct in this regime.  Examining the communities in
detail reveals that the method occasionally splits or merges communities.
It is possible that performance could be improved further by a less
simple-minded post-processing step for optimizing the likelihood.

\begin{figure}
\begin{center}
\includegraphics[width=8cm]{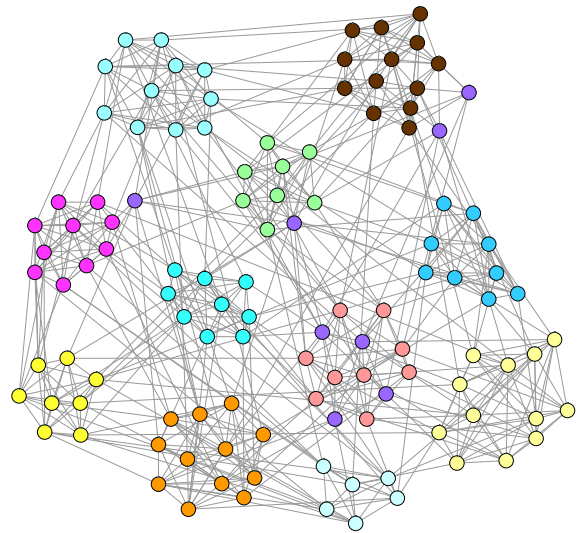}
\end{center}
\caption{Non-overlapping communities found in the US college football
  network of~\cite{GN02}.  The clusters of vertices represent the
  communities found by the algorithm, while the vertex colors represent the
  ``conferences'' into which the colleges are formally divided.  As we can
  see, the algorithm in this case extracts the known conference structure
  perfectly.  (The dark purple vertices represent independent colleges that
  belong to no conference.)}
\label{fig:football}
\end{figure}

We also give, in Fig.~\ref{fig:football}, an example of a test of the
method against a real-world network, in this case the much studied college
football network of Ref.~\cite{GN02}.  In this network the vertices
represent university teams in American football and the edges represent the
schedule of games for the year 2000 football season, two teams being
connected if they played a game.  It has been found in repeated analyses
that a clustering of this network into communities can retrieve the
organizational units of US college sports, called ``conferences,'' into
which universities are divided for the purposes of competition.  In 2000
there were 11 conferences among the Division~I-A teams that make up the
network, as well as 8 teams independent of any conference.  As
Fig.~\ref{fig:football} shows, every single team that belongs in a
conference is placed correctly by our algorithm.

\section{Conclusion}
In this paper we have described a method for detecting communities, either
overlapping or not, in undirected networks.  The method has a rigorous
mathematical foundation, being based on a probabilistic model of link
communities; is easy to implement, fast enough for networks of millions of
vertices; and gives results competitive with other algorithms.

Nonetheless, the method is not perfect.  Its main current drawback is that
it offers no criterion for determining the value of the parameter we
call~$K$, the number of communities in a network.  This is a perennial
problem for community detection methods of all kinds.  Some methods, such
as modularity maximization, do offer a solution to the problem, but that
solution is known to give biased answers or be inconsistent under at least
some circumstances~\cite{FB07,Bickel2009}.  More rigorous approaches such
as the Bayesian information criterion and the Akaike information criterion
are unfortunately not applicable here because many of the model parameters
are zero, putting them on the boundary of the parameter space, which
invalidates the assumptions made in deriving these criteria.

Another approach is to perform the calculations with a large value of~$K$
and regularize the parameters in a manner such that some communities
disappear, meaning that zero edges are associated with those communities.
For example, Psorakis~\etal~\cite{Psorakis2010}, in studies using their
matrix factorization algorithm, used priors that penalized their model for
including too many nonzero parameter values and hence created a balance
between numbers of communities and goodness of fit to the network data.
Unfortunately, the priors themselves contain undetermined parameters whose
values can influence the number of communities and hence the problem is not
completely solved by this approach.

We believe that statistical model selection methods applied to generative
models should in principle be able to find the number of communities in a
consistent and satisfactory manner.  We have performed some initial
experiments with such methods, and the quality of the results seems
promising, but the methods are at present too computationally demanding to
be applied to any but the smallest of networks.  It is an open problem
whether a reliable method can be developed that runs in reasonable time on
the large networks of interest to today's scientists.

\begin{acknowledgments}
  The authors thank Qiaozhu Mei, Cris Moore, and Lenka Zdeborova for useful
  conversations.  This work was funded in part by the National Science
  Foundation under grant DMS--0804778 and by the James S. McDonnell
  Foundation.
\end{acknowledgments}

\appendix
\section{Community detection and
statistical text analysis}
\label{app:plsa}
As mentioned in the main text, the generative model we use is the network
equivalent of a model used in text analysis called
probabilistic latent semantic analysis (PLSA)~\cite{Hofmann1999,
  Hofmann2001, Hofmann2004}, modified somewhat for the particular problem
we are addressing.  In this appendix, we describe PLSA and related methods
and models and their relationship to the community detection problem.

A classic problem in text analysis, which is addressed by the PLSA method,
is that of analyzing a ``corpus'' of text documents to find sets of words
that all (or mostly) occur in the same documents.  The assumption is that
these sets of words correspond to topics or themes that can be used to
group documents according to content.  The PLSA approach regards documents
as a so-called ``bag of words,'' meaning one considers only the number of
times each word occurs in a document and not the order in which words
occur.  (Also, one often considers only a subset of words of interest,
rather than all words that appear in the corpus.)

Mathematically a corpus of $D$ documents and $W$ words of interest is
represented by a matrix~$A$ having elements $A_{wd}$ equal to the number of
times word~$w$ appears in document~$d$.  To make the connection to
networks, this matrix can be thought of as the incidence matrix of a
weighted bipartite network having one set of vertices for the documents,
one for the words, and edges connecting words to the documents in which
they appear with weight equal to their frequency of occurrence.

In PLSA each word-document pair---an edge in the corresponding network
picture---is associated with an unobserved variable~$z$ which denotes one
of $K$ topical groups.  Each edge is assumed to be placed independently at
random in the bipartite graph, with the probability that an edge falls
between word~$w$ and document~$d$ being broken down in the form $\sum_z
P(w|z)P(d|z)P(z)$, where $P(z)$ is the probability that the edge belongs to
topic~$z$, $P(w|z)$ is the probability that an edge with topic~$z$ connects
to word~$w$, and $P(d|z)$ is the probability that an edge with topic~$z$
connects to document~$d$.  Note that, given the topic, the document and
word ends of each edge are placed independently.
(Hofmann~\cite{Hofmann1999} calls this parametrization a ``symmetric'' one,
meaning that the word and the document play equivalent roles
mathematically, but in the networks jargon this would not be considered a
symmetric formulation---the network is bipartite and the incidence matrix
is not symmetric, nor even, in general, square.)

An alternative description of the model, which is useful for actually
generating the incidence matrix and which corresponds with our formulation
of the equivalent network problem, is that each matrix element~$A_{wd}$
takes a random value drawn independently from a Poisson distribution with
mean $\sum_z P(w|z) P(d|z) \,\omega_z$.  In the language of networks, each
edge is placed with independent probability $\sum_z P(w|z)P(d|z)P(z)$,
where $P(z) = \omega_z/\sum_{z'} \omega_{z'}$.  In our work, where we focus
on one-mode networks and a symmetric adjacency matrix instead of an
incidence matrix, the parameter~$\omega_z$ is redundant and we omit it.

PLSA involves using the edge probability above to calculate a likelihood
for the entire word-document distribution, then maximizing with respect to
the unknown probabilities $P(w|z)$, $P(d|z)$, and~$P(z)$.  The resulting
probabilities give one a measure of how strongly each word or document is
associated with a particular topic~$z$, but since the topics are arbitrary,
this is effectively the same as simply grouping the words and documents
into ``communities.''  Alternatively, one can use the probabilities to
divide the edges of the bipartite graph among the topical groups, giving
the text equivalent of the ``link communities'' that are the focus of our
calculations.

A number of methods have been explored for maximizing the likelihood.
Mathematically the one most closely related to our approach is the
expectation-maximization (EM) algorithm of
Hoffman~\cite{Hofmann1999,Hofmann2001,Hofmann2004}, though the
correspondence is not exact.  Hofmann's work focuses solely on text
processing---the connection to networks was not made until later---and the
method cannot be translated directly for applications to standard
undirected one-mode networks.  It is possible to generalize the method to
directed one-mode networks in a fairly straightforward fashion, and one
might imagine that the undirected case could then be treated as a directed
network with directed edges running in both directions between every
connected pair of vertices.  Unfortunately, this approach does not work,
since in general it results in a posterior edge probability~$q_{ij}(z)$
which is asymmetric in $i$ and~$j$.  This asymmetry causes problems when we
want to associate a community with an undirected edge.  If $q_{ij} \ne
q_{ji}$, then the edge may be in one community when considered as an edge
from $i$ to~$j$ and a different community when considered as an edge from
$j$ to~$i$.

Rather than applying the standard PLSA model to network problems,
therefore, a better approach is to use a model that is intrinsically
symmetric from the outset, and this leads us to the formulation in this
paper.  This symmetric formulation and the corresponding EM algorithm have
not, to our knowledge, been used previously for community detection in
networks, but several other related approaches have, including ones based
on the techniques known as nonnegative matrix factorization
(NMF)~\cite{Lee99,Ding2008} and latent dirichlet allocation
(LDA)~\cite{Blei2003, Girolami2003}.  These formulations have similar goals
to ours, but are typically asymmetric (and hence unsuitable for undirected
networks) and use different algorithmic approaches for maximizing the
likelihood.  The NMF formulation is similar in style to an EM algorithm,
using an iterative maximization scheme, but the specific iteration
equations are different.  Several papers have recently proposed using NMF
to find overlapping communities~\cite{Zarei2009,Wang2010,Psorakis2010}, and
the work of Psorakis~\etal~\cite{Psorakis2010} in particular uses NMF with
the PLSA model, although again in an asymmetric formulation, and not
applied to link communities.

Recent work by Parkinnen~\etal~\cite{Parkinnen2009} and Gyenge~\etal\
\cite{Gyenge2010} does consider link communities, in an asymmetric
formulation, but uses algorithmic approaches that are different again.
Parkinnen~\etal~\cite{Parkinnen2009} use a model that attaches conjugate
priors to the parameters and then samples the posterior distribution of
link communities with a collapsed Gibbs sampler.

LDA~\cite{Blei2003,Girolami2003} offers an alternative but related approach
that also attaches priors to the parameters, but in a specific way that
relies on the asymmetric formulation of the model.  In~\cite{Henderson2009}
and~\cite{Zhang07}, LDA is adapted to networks by treating vertex-edge
pairs as analogous to word-document pairs and then associating communities
with the vertex-edge pairs.  This is an interesting approach but differs
substantially from the others discussed here, including our own, in which
vertex-vertex pairs (i.e.,~edges) are the quantity analogous to
word-document pairs.

Finally, in Appendix~\ref{app:relaxation} we show that our model can be
used to find nonoverlapping communities by viewing it as a relaxation of a
nonoverlapping stochastic blockmodel.  A corresponding relaxation has been
noted previously for a version of NMF and was shown to be related to
spectral clustering~\cite{Ding05,Ding2008a}.

\section{Results for running time}
\label{app:alg}
As discussed in Section~\ref{sec:implementation}, a naive implementation of
the EM equations gives an algorithm that is only moderately fast---not fast
enough for very large networks.  We described a more sophisticated
implementation that prunes unneeded variables from the iteration and
achieves significantly greater speed.  In this appendix we give a
comparison of the performance of the two versions of the algorithm on a set
of test networks.

The results are summarized in Table~\ref{tab:speedtable}, which gives the
CPU time in seconds taken to complete the overlapping community detection
calculation on a standard desktop computer (\textit{circa}~2011) for each
of the test networks.  In these tests we use 100 random initializations of
the variables and take as our final result the run that gives the highest
value of the log-likelihood.  For each network we give the results of three
different calculations: (1)~the calculation performed using the naive EM
algorithm; (2)~the calculation using the pruned algorithm with the
threshold parameter $\delta$ set to zero, meaning the algorithm gives
results identical to the naive algorithm except for numerical rounding; and
(3)~the calculation performed using the pruned algorithm with
$\delta=0.001$, which introduces an additional approximation that typically
results in a slightly poorer final value of the log-likelihood, but gives a
significant additional speed boost.

The largest network studied, which is a network of links in the online
community LiveJournal, is an exception to the pattern: for this network,
which contains over 40 million edges, we performed single runs only, with
one random initialization each, using the pruned algorithm with
$\delta=0.001$ and with $\delta=0$.  The run with $\delta=0.001$ took about
50~minutes to complete and the run with $\delta=0$ took about 11 hours.

\begin{table*}
\begin{centering}
\begin{footnotesize}
\setlength{\tabcolsep}{6pt}
\begin{tabular}{lrrr}
Running conditions & Time (s) & Iterations & Log-likelihood \\
\hline
\multicolumn{4}{c}{US air transportation, $n=709$, $m=3327$, $K=3$} \\
\hline
naive, $\delta = 0$    & 15.71 & 55719 & $-8924.58$ \\
fast, $\delta = 0$     & 14.67 & 55719 & $-8924.58$ \\
fast, $\delta = 0.001$ & 2.17  & 26063 & $-9074.21$  \\
\hline
\multicolumn{4}{c}{Network science collaborations \cite{Newman06c}, $n=379$, $m=914$, $K=3$} \\
\hline
naive, $\delta = 0$    & 0.93 & 13165 & $-3564.74$ \\
fast, $\delta = 0$     & 0.82 & 13165 & $-3564.74$ \\
fast, $\delta = 0.001$ & 0.13 & 10747 & $-3577.85$ \\
\hline
\multicolumn{4}{c}{Network science collaborations, $n=379$, $m=914$, $K=10$} \\
\hline
naive, $\delta = 0$    & 3.19 & 18246 & $-2602.15$ \\
fast, $\delta = 0$     & 3.15  & 18246 & $-2602.15$ \\
fast, $\delta = 0.001$ & 0.49  & 12933 & $-2611.96$ \\
\hline
\multicolumn{4}{c}{Network science collaborations, $n=379$, $m=914$, $K=20$} \\
\hline
naive, $\delta = 0$    & 6.16 & 19821 & $-2046.95$ \\
fast, $\delta = 0$     & 6.09 & 19821 & $-2046.95$ \\
fast, $\delta = 0.001$ & 0.94  & 14010 & $-2094.85$ \\
\end{tabular}
\hfill
\begin{tabular}{lrrr}
Running conditions & Time (s) & Iterations & Log-likelihood \\
\hline
\multicolumn{4}{c}{Political blogs \cite{AG05}, $n=1490$, $m=16\,778$, $K=2$} \\
\hline
naive, $\delta = 0$    & 11.42 & 13773 & $-48761.1$ \\
fast, $\delta = 0$     & 11.46 & 13773 & $-48761.1$ \\
fast, $\delta = 0.001$ & 4.14 & 13861 & $-48765.6$ \\
\hline
\multicolumn{4}{c}{Physics collaborations \cite{Newman01a}, $n=40\,421$, $m=175\,693$, $K=2$} \\
\hline
naive, $\delta = 0$    & 4339.57 & 424077 & $-1.367 \times 10^6$ \\
fast, $\delta = 0$     & 2557.91 & 424077 & $-1.367 \times 10^6$ \\
fast, $\delta = 0.001$ & 253.41  & 61665  & $-1.378 \times 10^6$ \\
\hline
\multicolumn{4}{c}{Amazon copurchasing \cite{Leskovec2003}, $n=403\,394$, $m=2\,443\,408$, $K=2$} \\
\hline
naive, $\delta = 0$    & 170646.9 & 1222937 & $-2.521 \times 10^7$ \\
fast, $\delta = 0$ & 105042.3 & 1222937 & $-2.521 \times 10^7$   \\
fast, $\delta = 0.001$ & 11635.0  & 120612 & $-2.538 \times 10^{7}$ \\
\hline
\multicolumn{4}{c}{LiveJournal \cite{Backstrom2006,Leskovec2008}, $n=4\,847\,571$, $m=42\,851\,237$, $K=2$} \\
\hline
fast, $\delta = 0$ & 39834.3 & 26163 & $-4.611 \times 10^8$ \\
fast, $\delta = 0.001$ & 3093.7 & 1389 & $-4.660 \times 10^8$ \\
\end{tabular}
\end{footnotesize}
\end{centering}
\caption{Example networks and running times for each of the three
  versions of the algorithm described in the text.  The designations
  ``fast'' and ``naive'' refer to the algorithm with and without pruning
  respectively.  ``Iterations'' refers to the total number of iterations
  for the entire run, not the average number for one random initialization.
  ``Time'' is similarly the total running time for all initializations.
  Directed networks were symmetrized for these tests.  All networks were run
  with 100 random initializations, except for the LiveJournal network,
  which was run with only one random initialization.}
\label{tab:speedtable}
\end{table*}

While the algorithm described is fast by comparison with most other
community detection methods, it is possible that its speed could be
improved further (or that the quality of the results could be improved
while keeping the speed the same).  Two potential improvements are
suggested by the text processing literature discussed in
Appendix~\ref{app:plsa}.  The first, from Hofmann~\cite{Hofmann2004}, is to
use the so-called tempered EM algorithm.  The second, from
Ding~\etal~\cite{Ding2008}, is to alternate between the EM algorithm and a
nonnegative matrix factorization algorithm, exploiting the fact that both
maximize the same objective function but in different ways.

\section{Nonoverlapping communities}
\label{app:relaxation}
In Section~\ref{sec:nonoverlap} we described a procedure for extracting
nonoverlapping community assignments from network data by first finding
overlapping ones and then assigning each vertex to the community to which
it belongs most strongly.  This procedure was presented as a heuristic
strategy for the nonoverlapping problem, but in this appendix we show how
it can be derived in a principled manner as an approximation method for
fitting the data to a degree-corrected stochastic blockmodel.

Methods have been proposed for discovering nonoverlapping communities in
networks by fitting to the class of models known as stochastic blockmodels.
As discussed in Ref.~\cite{Karrer2010SB}, it turns out to be crucial that
the blockmodel used incorporate knowledge of the degree sequence of the
network if it is to produce useful results, and this leads us to consider
the so-called degree-corrected blockmodel, which can be formulated as
follows.  We consider a network of~$n$ vertices, with each vertex belonging
to exactly one community.  The community assignments are represented by an
indicator variable~$S_{ir}$ which takes the value~1 if vertex $i$ belongs
to community $r$ and zero otherwise.  To generate the network, we place a
Poisson distributed number of edges between each pair of vertices~$i,j$,
such that the expected value of the adjacency matrix element is
$\theta_i\omega_{rs}\theta_j$ if vertex~$i$ belongs to group~$r$ and
vertex~$j$ belongs to group~$s$, where $\theta_i$ and $\omega_{rs}$ are
parameters of the model.  To put this another way, the mean value of the
adjacency matrix element is $\theta_i \bigl( \sum_{rs} S_{ir} \omega_{rs}
S_{js} \bigr) \theta_j$ for every vertex pair.  The normalization of the
parameters is arbitrary, since we can rescale all $\theta_i$ by the same
constant if we simultaneously rescale all~$\omega_{rs}$.  In our
calculations we fix the normalization so that the $\theta_i$ sum to unity
within each community: $\sum_i \theta_i S_{ir} = 1$ for all~$r$.

Now one can fit this model to an observed network by writing the
probability of generation of the network as a product of Poisson
probabilities for each (multi-)edge, then maximizing with respect to the
parameters $\theta_i$ and $\omega_{rs}$ and the community
assignments~$S_{ir}$.  Unfortunately, while the maximization with respect
to the continuous parameters~$\theta_i$ and $\omega_{rs}$ is a simple
matter of differentiation, the maximization with respect to the discrete
variables~$S_{ir}$ is much harder.  A common way around such problems is to
``relax'' the discrete variables, allowing them to take on continuous real
values, so that the optimization can be performed by differentiation.  In
the present case, we would allow the $S_{ir}$ to take on arbitrary
non-negative values, subject to the constraint that $\sum_r S_{ir} = 1$.
In effect, $S_{ir}$~now represents the fraction by which vertex~$i$ belongs
to group~$r$, with the constraint ensuring that the fractions add correctly
to~1.

With this relaxation, we can now absorb the parameters $\theta_i$ into
the~$S_{ir}$, defining $\theta_{ir}=\theta_i S_{ir}$ with $\sum_i
\theta_{ir} = 1$, and the mean number of edges between vertices $i$ and~$j$
becomes $\sum_{rs} \theta_{ir}\omega_{rs}\theta_{js}$.  This is an extended
form of the overlapping communities model studied in this paper,
generalized to include the extra $K\times K$ matrix~$\omega_{rs}$.  In the
language of link communities, this generalization gives us a model in which
the two ends of an edge can belong to different communities.  One can think
of each end of the edge as being colored with its own color, instead of the
whole edge taking only a single color.  If $\omega_{rs}$ is constrained to
be diagonal, then we recover the single-color version of the model again.

We can fit the general (nondiagonal) model to an observed network using an
expectation-maximization algorithm, just as before.  Defining a
probability~$q_{ij}(r,s)$ that an edge between $i$ and $j$ has colors $r$
and~$s$, the EM equations are now
\begin{equation}
q_{ij}(r,s) = \frac{\theta_{ir}\omega_{rs}\theta_{js}}{\sum_{rs}
  \theta_{ir} \omega_{rs} \theta_{js}},
\end{equation}
and
\begin{equation}
\theta_{ir} = \frac{\sum_{js} A_{ij} q_{ij}(r,s)}{\sum_{ijs}
  A_{ij}q_{ij}(r,s)},\qquad
\omega_{rs} = \sum_{ij}A_{ij}q_{ij}(r,s).
\end{equation}
By iterating these equations we can find a solution for the
parameters~$\theta_{ir}$.  But $\theta_{ir} = \theta_i S_{ir}$ and, summing
both sides over~$r$, we get $\sum_r \theta_{ir} = \theta_i$, since $\sum_r
S_{ir} = 1$.  Hence
\begin{equation}
S_{ir} = {\theta_{ir}\over\theta_i} = {\theta_{ir}\over\sum_r \theta_{ir}}.
\end{equation}
Thus we can calculate the values of $S_{ir}$ and once we have these we can
then reverse the relaxation of the model by rounding the values to zero or
one, which is equivalent to assigning each vertex~$i$ to the community~$r$
for which $S_{ir}$ is largest, or equivalently the community for
which~$\theta_{ir}$ is largest.

Thus the final algorithm for dividing the network is simply to iterate the
EM equations to convergence and then assign each vertex to the community
for which~$\theta_{ir}$ is largest.  This is precisely the algorithm that
we used in Section~\ref{sec:nonoverlap}, except that the model is
generalized to include the matrix~$\omega_{rs}$, where in our original
calculations this matrix was absent which is equivalent to assuming it to
be diagonal.  In our experiments, however, we have found that even when we
allow~$\omega_{rs}$ to be nondiagonal, the algorithm commonly chooses a
diagonal value anyway, which implies that the output of our original
algorithm and the generalized algorithm should be the same.  (We note that
in practice the diagonal version of the algorithm runs faster, and that
both are substantially faster than the vertex moving heuristic proposed for
the stochastic blockmodel in Ref.~\cite{Karrer2010SB}.)

It is entirely possible, however, that there could be networks with
interesting nondiagonal group structure that could be detected using the
more general model.  The model including the matrix~$\omega_{rs}$ can in
principle find disassortative community structure---structure in which
connections are less common within communities than between them---as well
as the better studied assortative structure.  For example, the model can
detect bipartite structure in networks, whereas the unadjusted model can
not.


\begin{thebibliography}{10}
\bibitem{note1} This is a special case of the general observation that the
  log of the average of any set of numbers is never less than the average
  of the log, since the logarithm is concave down.  If the numbers in
  question are $x_z/q_z$ and the average is taken with weights~$q_z$ this
  observation leads immediately to the inequality given.
\bibitem{note2} If the value of $\delta$ is greater than $1/K$, then it is
  possible inadvertently to prune all of the colors from a vertex, leaving
  it in no community at all.  To avoid this, we must choose $\delta<1/K$,
  which we have done in all the calculations presented here.
\end{thebibliography}
\end{document}